# An Agent-based Strategy for Deploying Analysis Models into Specification and Design for Distributed APS Systems

Luis Antonio de Santa-Eulalia[1], Sophie D'Amours[2] and Jean-Marc Frayret[3]

[1] Téluq, Université du Québec à Montréal
Québec City, Québec, Canada

[2] Université Laval
Québec City, Québec, Canada

[3] École Polytechnique de Montréal
Montréal, Québec, Canada

## Abstract

Despite the extensive use of the agent technology in the Supply Chain Management field, its integration with Advanced Planning and Scheduling (APS) tools still represents a promising field with several open research questions. Specifically, the literature falls short in providing an integrated framework to analyze, specify, design and implement simulation experiments covering the whole simulation cycle. Thus, this paper proposes an agent-based strategy to convert the 'analysis' models into 'specification' and 'design' models combining two existing methodologies proposed in the literature. The first one is a recent and unique approach dedicated to the 'analysis' of agent-based APS systems. The second one is a well-established methodological framework to 'specify' and 'design' agent-based supply chain systems. The proposed conversion strategy is original and is the first one allowing simulation analysts to integrate the whole simulation development process in the domain of distributed APS.

*Keywords:* *Advanced Planning and Scheduling (APS), Agent-Based Simulation, Methodological Framework, Analysis, Specification and Design, FAMASS.*

## 1. Introduction

Advanced Planning and Scheduling (APS) systems comprise a set of techniques for the supply chain planning over short, intermediate, and long-term time periods. They employ advanced mathematical algorithms or logic to perform optimization or simulation on finite capacity scheduling, sourcing, capital planning, resource planning, forecasting, demand management, and other. APS simultaneously considers a range of constraints and business rules to provide real-time planning and scheduling, decision support, available-to-promise, and capable-to-promise capabilities. In addition, these systems often generate and evaluate multiple 'what-if' scenarios [1].

The use of these sophisticated optimization approaches in complex real-life supply chain situations has recently become possible mainly due to the increased computing power of companies [2].

Despite the contribution of APS systems to the supply chain planning domain, some criticism exists in this area [3]. Traditional APSs are basically monolithic systems that cannot model and take into account the complex everyday interactions and information exchanges between partners. For example, APS systems are deficient in handling sophisticated interaction mechanisms that allow the implementation of delegation and coordination approaches, which are methodologies based on negotiation, and cooperation strategies [4, 5]. As a result, the focus on relationships in a multi-tier environment has only recently been claimed by the APS community [6].

To cope with this problem, recent advances in supply chain planning have arisen in the area of agent technology. This technology is able to capture the distributed nature of supply chain entities (e.g. customers, manufacturers, logistics operators etc.) and mimic their business behaviours (e.g. making advanced production decisions and negotiating with other supply chain members), thus supporting their collaborative planning process. Because of these abilities, among several others described in the literature, agent-based supply chain systems have great potential for simulating complex and realistic scenarios [7, 4; 9, 10, 11]. Distributed APS systems employing agent-based technology are referred to in this paper as distributed APS systems [12].

Distributed APS systems are normally developed through the use of modelling and simulation frameworks and, usually, these frameworks provide principles, steps, methods and tools for creating a model. They help people understand the simulation problem to be modelled and translate it into a computing model normally used in simulation experiments in the supply chain planning area.

In order to create such models, these frameworks guide simulation modellers through one or several development steps [13]. The first modelling step is *analysis*, where one defines an abstract description of the modelled supply chain planning system containing functional and non-functional requirements. Next, during *specification,* the information derived from the analysis is translated into a formal model. As the analysis phase does not necessarily allow obtaining a formal model, the specification examines the analysis requirements and builds a model based on a formal approach. After, in the *design* phase one creates a data-processing model that describes the specification model in more detail. In the case of an agent-based system, design models are close to how agents operate. Finally, during *implementation,* the design model is translated into a specific software platform or it is programmed [13].

The problem behind these modelling frameworks is that normally simulation systems are implemented as directed by pre-stated requirements with little explicit focus on system analysis, specification, design and implementation in an integrated manner [14]. According to a recent literature review [15], to the best of our knowledge there are no integrated modelling approaches covering the whole developed process in this area. Moreover, there is one unique 'analysis' modelling, the FAMASS (*F*ORAC *A*rchitecture for *M*odelling *Ag*ent-based *S*imulations for *S*upply chain planning) framework, dedicated to the distributed APS domain, and which was proposed by us recently [21, 22, 23].

Despite its contribution to the literature, FAMASS is limited to the identification and mapping of functional requirements of distributed APS simulations, i.e. the 'analysis' phase only. If the simulation analysts desire to go further in the modelling process, they have to employ another 'specification' and 'design' methodology. This can be laborious, since analysts need to thoroughly master FAMASS and another methodology.

In order to facilitate FAMASS analysts in converting their analysis models into specification and design models, this paper proposes an agent-based deployment strategy. This strategy enlarges the FAMASS scope to the other modelling phases, thus covering the entire modelling cycle. By doing so, analysis can go smoother and quicker through this cycle.

To do so, we were inspired by the specification and design principles of the Labarthe et al. [9] framework, a recent and largely cited development in the field of methodological agent-oriented framework for supply chain management simulation. Since the focus of this framework is on supply chain management as a general concept (and not specialized in APS systems), we had to perform some minor adaptations to this approach. Despite these adaptations, the main ideas of Labarthe et al. [9] are explicitly considered in the deployment strategy. The Labarthe et al. framework is adopted here because it covers the specification and design phases properly at the business and agent levels, just as FAMASS does, which facilitates the deployment process.

This deployment strategy demonstrates that the analysis phase of FAMASS can be integrated with other existing approaches specialized in specification and design modelling. Furthermore, it allows us to avoid the research effort needed to develop a totally new specification and design methodology for the domain, although it would be suitable (and even desirable) for future research initiatives.

This paper is organized as follows: a literature review in modelling and simulation for distributed APS systems is presented in Section 2. Section 3 introduces the FAMASS approach, while Section 4 summarizes the Labarthe et al. [9] framework. Next, the deployment process is explained in Section 5. Finally, Section 6 outlines some final remarks and suggests future work.

## 2. Modelling and Simulation Frameworks for distributed APS

The use of agent technology in Supply Chain Management is a fruitful field. From the inaugural work of Fox et al. [16] until today, a large variety of works have appeared to propose different ways of encapsulating supply chain entities and performing simulation experiments.

Two types of modelling approaches can be identified in the literature. The first type proposes generic approaches for modelling agent-based supply chain systems in general terms, while the second type proposes a modelling framework that specifically takes into consideration Advanced Planning and Scheduling (APS) tools when planning, i.e. the incorporate optimization procedures or finite capacity planning models when performing supply chain planning. APS systems emerged in the last decade to provide a suite of planning and scheduling modules for the firm's internal supply chain, from the raw materials source to the consumers and covering decisions ranging from the strategic to the operational level [17].

In the first type of approach (general agent-based models), examples of relevant contributions include Labarthe et al. [9], Van der Zee, and Van der Vorst [18], MaMA-S [13]. One of the most cited works in the domain is Labarthe et al. [9], which propose a methodological framework for modelling customer-centric supply chains in the context of mass customization. They define a conceptual model for supply chain modelling and show how the multi-agent system can be implemented using predefined agent platforms. Van der Zee and Van der Vorst [18] propose an agent framework derived from an object-oriented approach to explicitly model control structures of supply chains. MaMA-S [13] provides a multi-agent methodology for a distributed industrial system, which is divided into five main phases and two support phases. The authors propose formal methods for the specification, design and implementation phases, but the analysis phase is not tackled by them.

This second type of modelling approach provides more sophisticated models of supply chains by incorporating Advanced Planning and Scheduling routines [12]. These approaches, sometimes called d-APS systems (for distributed APS), are composed of semi-autonomous APS tools, each dedicated to a specialized planning area and that can act together in a collaborative manner employing sophisticated interaction schemas.

Examples of this kind of work are Egri et al. [19], Lendermann et al. [20] and Swaminathan et al. [11]. Egri et al. [19] is a Gaia-based approach for modelling advanced distributed supply chain planning for mass customization. They develop a model for representing roles and interactions of agents based on the SCOR (Supply-Chain Operations Reference) model. Lendermann et al. [20] developed an approach to couple discrete-event simulation and APS for collaborative supply chain optimization, based on the HLA (High Level Architecture) technology for distributed simulation synchronization. Swaminathan et al. [11] provide a supply chain modelling framework containing a library of modular and reusable software components, which represents different kinds of supply chain agents, their constituent control elements and their interaction protocols.

These simulation and modelling approaches have greatly contributed to the domain, however, in spite of these advances, there exists a relevant gap in this field related to the initial developing step of such simulation systems, the analysis phase [12]. Most of the researched works in the literature suggest approaches for specification and design, and some for implementation, but the analysis phase is not explicitly treated [12, 13, 14, 21]. Most of these works suppose that the analysis phase furnishes the necessary information and concentrate their discussions on further phases, mainly specification and design. The first work dedicated to the analysis of distributed APS systems using the agent-based paradigm is FAMASS [21]. Despite its contribution to the agent-based modelling of distributed APS systems, FAMASS does not cover the specification and design phases of the development process. This is an interesting research gap in the literature. Section 3 details the FAMASS approach for the analysis phase, while Section 4 presents a frequently cited method for specification and design of agent-based supply chain systems from Labarthe et al. [9]. Next, Section 5 combines these two approaches in order to create a deployment strategy to translate analysis models into specification and design.

# 3. The FAMASS Approach

The FAMASS (*F*ORAC *A*rchitecture for *M*odelling *A*gent-based *S*imulation for *S*upply chain planning) is the first and unique modelling approach dedicated to the analysis phase of distributed APS simulations [21, 22, 23]. This approach was recently tested in Santa-Eulalia et al. [24].

It is organized into two abstraction levels: Supply chain: refers to the supply chain planning problem, i.e. the business viewpoint; Agent: the supply chain domain problem is translated into an agent-based view (Figure 1).

At these two abstraction levels, four modelling approaches are proposed, namely the General Problem Analysis (GPA), the Distributed Planning Analysis (DPA), the Social Agent Organization Analysis (SAOA) and the Individual Agent Organization Analysis (IAOA), as schematized in Fig. 1.

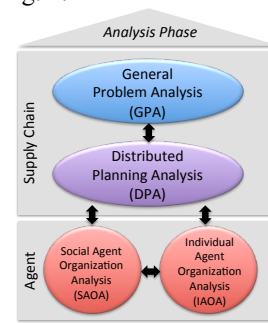

Fig. 1: Four main modelling approaches proposed for analysis of supply chain and agent levels [23].

These four modelling approaches are explained in the following subsections.

### 3.1 General Problem Analysis (GPA)

GPA is the first modelling effort where simulation analysts have to think about the simulation problems. The GPA is based on Santa-Eulalia et al. [12], in which a discussion about the simulation objective and the problem structure is provided.

Basically, the GPA proposes that the simulation analysis has to take two main aspects into consideration: general aspects and experimental aspects. General aspects represent macro definitions of the simulation problem, including the object and environment to be simulated, the simulation questions, hypotheses and objectives. Experimental aspects are related to the design of experiments, where one defines the factors, uncertainties and key performance indicators of the simulation.

These elements refer to the general definition of the simulation problem, according to what is desired to be studied, and it will guide the whole development process.

This general definition is then organized through some formalisms from SysML (Systems Modeling Language) [25]. In this case, some Requirements Diagrams help the analysts organize the GPA. An example of how this can be done is provided in [23].

### 3.2 Distributed Planning Analysis (DPA)

The DPA identifies what the desired supply chain planning entities are, as well as their roles. These entities are identified according to their mission in the supply chain and their planning functions at different decision levels.

To identify the main supply chain planning entities, FAMASS employs the concepts of supply chain integration proposed by Shapiro [26]. The author states that supply chain management refers to integrated planning relying on three basic dimensions: i) *Intertemporal dimension*: refers to different decision levels, i.e. strategic, tactical and operational decision levels; ii) *Functional dimension*: stands for different planning functions in a supply chain, which can be related to procurement, manufacturing, distribution and sales; iii) *Spatial dimension*: refers to the fact that supply chains are composed of geographically dispersed units of analysis.

This gives rise to the notion of a Supply Chain Block. A Supply Chain Block can be defined as a supply chain planning entity, which is a functional unit capable of: performing part of the supply chain planning decisions or their totality; or performing the execution of the supply chain decisions (part of them or their totality). These entities have a certain degree of autonomy and are able to interact with each other. Possible Supply Chain Blocks for covering the integrated supply chain planning dimensions are proposed in the framework of Fig. 2, which is called the supply chain planning cube.

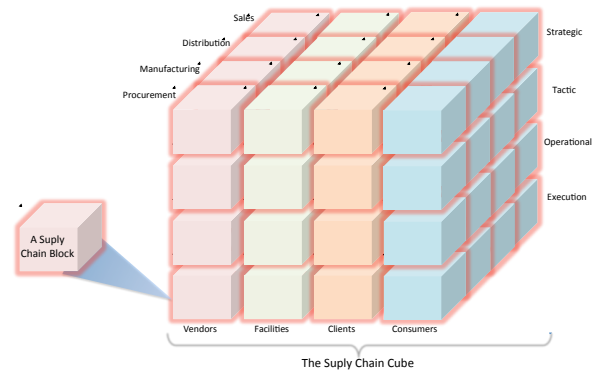

Fig. 2: Supply Chain Planning Cube [23].

A vertical slice of the supply chain planning cube for one spatial unit of analysis (e.g. facilities) is similar to the planning matrix proposed by Meyr and Stadtler [27], except for the execution level. The supply chain planning cube is an evolution of the planning matrix, due to the fact that it represents the possibility of collaboration among different traditional APS systems. It also includes execution entities.

Based on the supply chain cube, one has to perform requirements determination for the simulation aspects. This cube serves as a metamodel to help simulation analysts identify their simulation requirements. For example, the analysts decide which kind of Supply Chain Blocks will be needed in their simulation experiments, providing the basic architectural aspects of the simulation system. Then, their requirements are organized through UML-based use cases and requirements diagrams from SyML. An example of the DPA is provided in Santa-Eulalia et al. [23].

### 3.3 Social Agent Organization Analysis (SAOA)

So far, the concept of Supply Chain Block has been used to represent entities responsible for part of the supply chain planning. Together they compose a population of entities interacting with each other, having a collective co-existence within the planning system. When these entities incorporate attitudes, orientations and behaviours comprising the interests, needs or intentions of other Supply Chain Blocks, they can be seen as social entities. They can exhibit complex actions that take into account the collectivity. A way to represent social entities is to model them as agents, thus creating multi-agent societies.

The general logic indicated that a Supply Chain Block can be directly translated into agents by adding agent abilities to them. This is based on the agentification definition of Shen et al. [28], who explain that the agentification process can be functional-based (i.e. white Supply Chain Block) or physical-based (i.e. gray Supply Chain Block).

However, in some situations a Supply Chain Block can be transformed into more than one agent, for example when specialization is required, in which case a planning agent can be specialized according to certain generic responsibility orientations, such as products, processors, processes or projects, to obtain faster or more precise responses for certain given situations. In other situations, apart from agents proceeding from the supply chain planning cube, different intermediary agents can be created to perform activities related to, e.g. the coordination of the agents' society. In addition, the agentification process can also include the representation of information sources, interfaces and other services.

The importance of this discussion relies on the notion that agentification is the basis for two mutually dependent aspects in agent-based systems which define the metamodel for the SAOA:

- *Social structures*: represent the agent system architecture [24] characterizing the blueprint of relationships, giving a high level view of how groups solve problems and the role each agent plays within the structure. There are diverse types of social structures, such as hierarchical, federated and autonomous.

- *Social protocols*: are agents' abilities concerning social aspects, normally related to cooperation principles (i.e. agents have to cooperate in order to plan the entire supply chain). Diverse abilities can be considered, like communication, grouping and multiplication, coordination, collaboration by sharing tasks and resources and conflict resolution through negotiation and arbitration.

Different social structures and protocols are provided in Santa-Eulalia [22].

Similar to the DPA, these two aspects of the SAOA serve as a metamodel to help simulation analysis identify their requirements for the simulation model. For example, different social protocols can be tested in the simulation. Then, requirements can be organized through agent-based use cases from AUML (Agent Unified Modelling Language) and requirements diagrams from SysML. An example of the SAOA is provided in Santa-Eulalia et al. [23].

### 3.4 Individual Agent Organization Analysis (IAOA)

As mentioned by Ferber [29], the task of assigning roles to every individual agent is normally regarded as the last phase in constructing an organization. The logic is that as soon as one knows what the functions to be assigned are, one defines individual specializations. These local assignments influence social protocols functioning inside their respective social structures. In addition, it also influences the local performance of the supply chain planning entities. This is the main idea of the IAOA.

At the individual level, agents can be organized according to different internal architectures but there is little consensus on how to conceive the internal architectures of agents [30] in the literature. In order to cope with this, the metamodel for the IAOA proposes that whatever the state of mind of an agent is (cognitive, reactive or hybrid), and whatever internal architecture an agent employs, an agent can be described simply according to its 'abilities'. This is the central point when performing simulation. An 'ability' can be defined as the quality of being able to perform an action, or facilitate the action's accomplishment. 'Abilities' allow for the implementation of actions and the determination of the system's behaviour, as well as the determination of its related performance.

Based on this notion, the metamodel defines two elements:

- *The Response Space*: stands for a collection of general abilities available for the agents, including very simple reactive abilities or sophisticated cognitive ones. For example, one agent can have a simple ability to monitor the inventory levels of the supply chain, or a complex ability to perform production planning employing an optimization method.

- *Capacity to Produce an Adapted Response*: represents the aptitude to choose which abilities have to be transformed into actions at a given time to respond to a given situation. This capacity can vary from elementary to complex. The simplest possible capacity is related to a reactive 'if-then' mechanism, where no cognition is necessary. For example, if the inventory level drops to a given threshold, the agent uses its procurement ability to start a procurement action. As the agent becomes more intelligent, more complex responses can be made for some given situations. For example, the linear "if-then" logic can be substituted by more complex approaches based on action optimization and learning.

Based on these two elements of the metamodel, one can carry out requirements determination for the simulation model, selecting the desired requirements in terms of agents' abilities. Similar to the SAOA, the IAOA's requirements are organized through agent-based use cases from AUML and requirements diagrams from SysML [23].

FAMASS is detailed in Santa-Eulalia et al. [21, 22, 23]. An application of this approach is presented in Santa-Eulalia et al. [24].

## 4. Labarthe et al.'s Methodological Framework

The Labarthe et al. [9] framework is schematized in Fig. 3 and is briefly described afterwards.

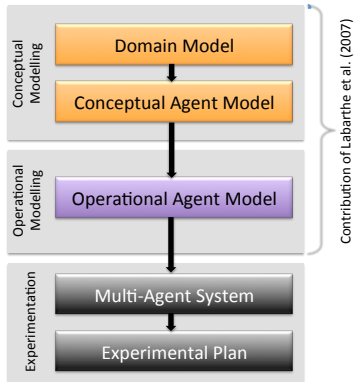

Fig. 3: Summarizing the Labarthes et al. [9] framework.

The authors propose the modelling steps indicated in Fig. 3. Their contribution corresponds to two abstraction levels: conceptual modelling and operational modelling. Conceptual modelling is performed in two steps, the Domain Model and the Conceptual Agent Model.

### 4.1 Domain Model (DM)

The Domain Model (DM) creates an abstraction of the supply chain. Inspired from the NetMAN approach [31, 32], Labarthe et al. [9] create two sub-models: a Structural Model and a Dynamic Model.

The Structural Model, which is based on responsibility networks [33], defines the structure of the supply chain, i.e. its 'actors' and their related responsibilities, and it also depicts the material flows among all 'actors'. The Dynamic Model complements the Structural Model by defining the behaviour of each 'actor' and its related interaction modes.

### 4.2 Conceptual Agent Model (CAM)

The Conceptual Agent Model (CAM) remodels the Domain Model guided by the agentification process. From the Structural and Dynamic models, a unique agent model is created. A Conceptual Agent Model specifies the 'agents', the 'objects' transacted between them and the nature of the agent's interactions ('physical interactions' and 'informational interactions'). In this case, each 'actor' specified in the Structural Model produces a specific agent. Also, any activity of an actor generates a specific agent in close interaction with the agent associated to the actor concerned, which is regrouped in the same partition. In addition, any exchange of information from the Dynamic Model generates a message-based informational interaction; and any material flow from the dynamic model leads to a physical type interaction.

After, at the Operational Level, Labarthe et al. [9] proposes the Operational Agent Model (OAM).

### 4.3 Operational Agent Model (OAM)

The Operational Agent Model (OAM) is based on the Conceptual Agent Model, and it aims to build a computer model of the studied supply chain which will be later implemented on a simulation platform. First, the Operational Agent models the software architecture (at the social level). Next, it models the internal agent architecture (individual level), dealing with knowledge, behaviours and interactions among agents.

After creating the Domain Model, the Conceptual Agent Model and the Operational Agent Model, a Multi-Agent System is implemented at the Exploitation level and a set of Experimental Plans supports the realization of simulation experiments (the black modelling approaches shown in Fig. 3). The author illustrated the Exploitation level though the implementation of a case study in a simulation environment.

This is only a summarized review of Labarthe et al. [9]'s work. For further details about this framework and its applications, the reader is referred to Labarthe et al. [9, 35, 36] and Labarthe [34].

## 5. The Deployment Process

As explained in the introduction, the original framework of Labarthe et al. [9] had to be slightly adapted to be suitable to the distributed APS domain.

The first adaptation occurs at the Domain Modelling. The main reason for not strictly employing the Labarthe et al. [9] Domain Model is because it is based on the responsibility network [33], which uses the definition of centre, i.e. a business entity – a decisional one – linked at the physical level by material flow. Centres do not correspond exactly to our semi-autonomous units, the Supply Chain Blocks (defined in subsection 3.2), which are based on the supply chain cube. We adapted the Labarthe et al. [9] model and thus proposed a modelling approach where the 'centres' are substituted by Supply Chain Blocks.

Another relevant difference refers to the fact that we separate the Operating System (i.e. the Execution layer) and the Decision System (i.e. the Strategic, Tactical and Operational layers) in the Domain Model, which is not done in the Labarthe et al. [9] Domain Model. They distinguish these two layers later, in the Operational Agent Model. We decided to separate them earlier because both systems have to be identified in regard to the supply chain cube introduced in subsection 3.2. If we did not consider entities of the Operating System at this step, the Domain Model would be incomplete for a distributed APS, according to the definition of the supply chain cube.

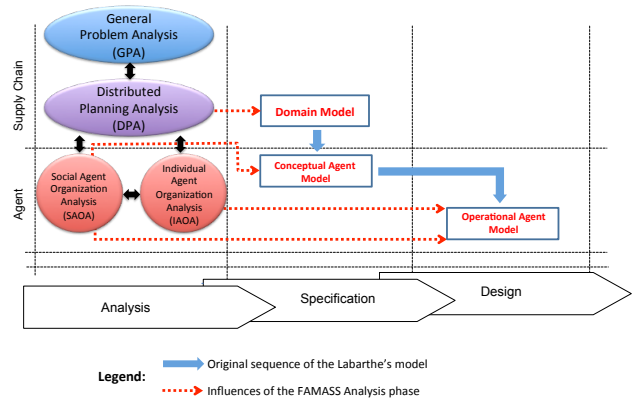

Fig. 4: Deploying process.

Fig. 4 depicts the general idea of the deployment process. From the analysis phase, the Distributed Planning Analysis models are the basis for the creation of the Domain Model. The Domain Model represents the supply chain under study and how advanced planning decisions are articulated. Next, the Conceptual Agent Model is naturally created from the Domain Model, but the Social Agent Organization Analysis is also used as an important reference. The Social Agent Organization Analysis provides the Social Structures for the Conceptual Agent Model and it reflects the agentification process used during the Social Agent Organization Analysis. Finally, the Operational Agent Model is created from the Conceptual Agent Model. However, relevant information about social protocols requirements comes from the Social Agent Organization Analysis, while requirements concerning the agents' abilities come from the Internal Agent Organization Analysis.

It is interesting to note, in Fig. 4, that the Domain Model and Conceptual Agent Model roughly correspond to the specification phase, while the Operational Agent Model can be considered equivalent to the design phase. The Domain Model and Conceptual Agent Model are the first formal models to describe the supply chain and the agent domain. The Operational Agent Model is closely related to how agents operate.

To sum up, FAMASS proposes a set of abstract notions for distributed APS systems, while Labarthe et al. [9] provide a formal and detailed description of how the system should work.

The following subsection discusses the Domain Model generation.

## 5.1 Domain Model (DM)

The objective of the Domain Model is to identify what is to be modelled in the supply chain. As seen in Fig. 4, the Distributed Problem Analysis (DPA) can be translated directly into the Domain Model.

Table 1 and Table 2 provide a translation strategy to create FAMASS Structural and Dynamic Models based on Labarthe et al. [9].

Table 1: Structural Model.

| Element | Labarthe et al. | FAMASS Counterpart |
|---|---|---|
| Central elements | **Main element**: a network of Centres [33] (roles and responsibilities) and their interactions.<br><br>**Roles**: Processor, producer, assembler, fulfiller, distributor, retailer, transporters, customer. Roles define the nature of the responsibility set.<br><br>**Responsibilities**: examples, packing, grouping, sales, etc.<br><br>**Organizational level**: supply chain, enterprise, business unit, cells, resources. | **Main element**: a network of Supply Chain Blocks and their interactions (interactions are simple representations of Supply Chain Block's relations). A Supply Chain Block is used instead of centres.<br><br>**Roles**: From the "spatial" axis of the supply chain cube (subsection 3.2), we identify the Supply Chain Blocks and their roles: vendors, facilities, clients and consumers.<br><br>**Responsibilities**: one can identify responsibilities from the 'functional' axis of the supply chain cube (subsection 3.2): procurement, manufacturing, distribution and sales.<br><br>**Organizational levels**: strategic, tactic, operational, execution for vendors, facilities, clients and consumers (i.e., the intertemporal axis). |
| Modelling formalism | Responsibility networks of Montreuil and Lefrançois [33]. | Class diagrams and class tables (from AUML – Agent Unified Modelling Language). The concept is the same as for responsibility network, but it is represented using AUML formalisms. Centres are classes; roles are roles in each class; responsibilities are operations in each class; organizational levels are stereotypes of the classes; business processes are operations in each class. |
| Modelling process | Identify decision elements of the supply chain and the physical interactions among them. | We identify the elements from the execution and decision systems and we add only the physical interactions. Informational interactions are added in the dynamic model (later on in the modelling process). |

Table 2: Dynamic Model.

| Element | Labarthe et al. | FAMASS Counterpart |
|---------|-----------------|--------------------|
| **Central elements** | Describes (in time) the system behaviour and the elements that compose it. Uses the responsibility network to recognize [33] the coordination modes by identifying the physical and informational relations used according to the environmental stimulus. | Describes the same elements, but with the possibility to add more information based on different experimental definitions, i.e. different configurations of the Supply Chain Blocks, and different performance indicators and uncertainties. |
| **Modelling formalism** | NetMan [31, 32] approach plus a representation of the decoupling point position. The decoupling point position is mentioned here because it is an important issue in the Labarthe et al. [9] framework. | Class diagrams and class tables (AUML). All flows are represented by arrows. The decoupling point is represented in the class name. Centre models are represented by arrows as well. Stock holding (raw material, work-in-process or final products) is represented in the operations of each class. |
| **Modelling process** | Apart from the physical flow identified previously, the modelling process describes the informational flow exchanged according to the dynamics of the environment.<br><br>Four informational flow types for coordination are identified: i) needs expression; ii) offers expression; iii) information about coordination; and iv) information sharing by models exchanges. In addition, the decoupling point is positioned and inventories are mapped (raw material, work-in-process and final product).<br><br>It identifies two models (for models exchange): the network model and the centre model. | The same flows are identified, as well as inventory positions and decoupling point position. They are described in the class tables. |

The most important difference between Labarthe et al. [9] and FAMASS is the use of centre for the former and the use of Supply Chain Block for the latter. Supply Chain Block is used instead of centres in FAMASS because decision entities are central elements. Labarthe [34, p.119] explains that a centre represents a decision process, but centre definitions are closely associated to physical entities of the execution system, i.e. there is a direct relation between a centre and an entity of the execution system. Later in the Labarthe et al. [9] modelling process, the decision system is introduced more formally in the Operational Agent Model. We separate the decision system from the execution system in the Domain Model, since we know that they are relevant for experimental definitions in distributed APS systems. Another difference is related to the fact that we employ a unique modelling formalism based on an AUML approach, coherent with the analysis phase of FAMASS, which employs only UML-inspired formalisms.

The next sub-section transforms the Domain Model into a Conceptual Agent Model.

## 5.2 Conceptual Agent Model (CAM)

The Conceptual Agent Model represents the agentification process of the Labarthe et al. [9] approach. The agentification process defines the agent society based on the Domain Model, i.e. which agents are created from the centres (in our case, Supply Chain Block) and how they are organized. Labarthe et al. [9] propose rules for creating agents (i.e., each centre becomes an actor-agent and each centre activity becomes an activity-agent). As discussed before, FAMASS converts each Supply Chain Block into an agent. It also verifies whether some agents are extinguished (e.g. merged with another agent) or whether new agents are introduced (e.g. a mediator). This information is obtained during the Social Agent Organization Analysis (SAOA).

As indicated in Fig. 4, the Conceptual Agent Model is generated from the Domain Model and the SAOA (in this case, the social structures). Using Labarthe et al. [9] rules, the Domain Model provides the basic classes' definition and, using the SAOA, it can be verified if new agent classes are derived from the Domain Model and if different social structures have to be tested and considered in the Conceptual Agent Model. Social Protocols from SAOA are not used in Conceptual Agent Modelling.

The Strategy for creating a Conceptual Agent Model is shown in Table 3.

Table 3: Conceptual Agent Models.

| Element | Labarthe et al. | FAMASS Counterpart |
|---|---|---|
| Central elements | **Actor-agent**: centre.<br><br>**Activity-agent**: represents a process of transformation, distribution, or stock keeping.<br><br>**Object**: products.<br><br>**Informational interaction**: same as in Domain Model.<br><br>**Physical interaction**: same as in Domain Model. | **Actor-agent**: agents representing an organizational unit of the supply chain (i.e. vendors, facilities, clients or customers), related to the 'spatial' axis. Actor-agents group several other agents, the activity-agents.<br><br>**Activity-agent**: agents from the decision system, representing the processes of procurement, manufacturing distribution and sales. These agents are at three different decision levels and they are related to the 'functional' axis.<br><br>**Objects**: defined products. This is the first time products are specified.<br><br>**Information interactions**: they come from the Domain Model.<br><br>**Physical Interactions**: they come from the Domain Model. |
| Modelling formalism | A graphical modelling formalism [34] that models the two types of agents and their interactions. The CAM model is derived from the DM model. | Adapted class diagrams, tables and package diagrams. The adaptation of the class diagrams refers to the insertion of objects (products), represented by simple square boxes in the link between two classes. |
| Modelling process | **1. From centre to actor-agent**: each centre creates an actor-agent.<br><br>**2. Physical interactions between actor-agents**: physical flow is specified by an arrow linking agents and indicating their respective exchanged objects.<br><br>**3. Informational interactions between actor-agents**: similar to 2, but for information flow.<br><br>**4. Organizational frontiers definition**: establishes the organization frontiers for the actor-agents and places the physical flows between the organizations.<br><br>**5. Definition of the activity-agents**: each activity of a centre is transformed into an activity-agent.<br><br>**6. Physical interactions between activity-agents**: specify the physical flow between the activity-agents and their related objects exchanged.<br><br>**7. Informational interactions between activity-agents**: same as 6, plus the interaction between actor-agents and activity-agents. | **Similar process, with the following differences:**<br>- Actor-agents and activity-agents: in the classes, use role definitions to indicate if it is an actor-agent or an activity-agent;<br>- Interactions: links between classes. |

It is important to note that an actor-agent coordinates a population of other activity-agents in the Labarthe et al. [9] approach. In the case of FAMASS, we decided to use the notion of actor-agent only as an aggregation of agents inside the same organization using a package diagram.

The next sub-section transforms the Conceptual Agent Model into an Operational Agent Model.

## 5.3 Operational Agent Model (OAM)

According to Labarthe [34], the OAM represents implementable models. These models involve a choice between two different agent architectures, i.e. the cognitive and the reactive architectures. We believe that most of the time it is not possible to completely distinguish cognitive agents from deliberative agents, meaning that normally agents can be seen as a hybrid state within the cognitive-reactive continuum. In Labarthe et al. [9]'s work, agents from the decision system assume a cognitive agent architecture, composing a cognitive agent society. Based on this society, the author then creates a reactive society responsible for the transformation process (execution system), linked with the cognitive society.

As we believe that the agents from the decision system can also assume reactive behaviours (see subsection 3.2), we prefer not to use this agent architecture notation for the Operational Agent Model. Instead, we create two societies (decision agents and execution agents) from the Conceptual Agent Model and start to define all agents' behaviours and agents' protocols in detail, as done by Labarthe et al. [9], which is not contradictory to Labarthe et al.'s [9] work. As explained before, instead of separating into decision and execution societies at the Operational Agent Model, our approach does it at the beginning of the specification phase, i.e. at the Domain Model.

In sum, our Operational Agent Model is generated from the Conceptual Agent Model, the Social Agent Organization Analysis and the Internal Agent Organization Analysis, as illustrated in Fig. 5.

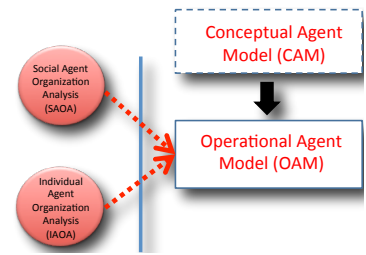

Fig. 5: Creating an Operational Agent Model.

From the Conceptual Agent Model we represent two societies, the decision agents and the execution agents. This is the starting point of the Operational Agent Model. After, we obtain requirements about agent protocols from the Social Agent Organization Analysis, and we obtain requirements about agent abilities from the Internal Agent Organization Analysis.

Table 4 summarizes the deployment strategy for the Operational Agent Model.

Table 4: Operational Agent Models.

| Element | Labarthe et al. | FAMASS Counterpart |
|---|---|---|
| **Central elements** | **Multi-agent system architecture**: a cognitive and a reactive agent society are represented. A cognitive agent, together with its corresponding reactive agent, form the 'agent-actor'. It is a generic architecture to represent entities capable of taking their own decisions and acting accordingly.<br><br>**Specification of the software agent**: knowledge, behaviour and interactions of each agent are defined. For the behaviours, the following entities are defined: a) external event: concerning the communication aspect with external entities of the multi-agent system; b) internal event: concerning internal activities of an agent; c) passive state: waiting state; d) active state, being an elementary action or a composite action. | **Multi-agent system architecture**: cognitive agents are seen as decision agents (from the decision system); reactive agents are represented by execution agents (from the execution system).<br><br>**Specification of the software agent**: same elements, i.e. knowledge, behaviour and interactions. |
| **Modelling formalism** | For the multi-agent system architecture, Labarthe [34] proposes his own graphical modelling formalism. For the specification of the software agent for cognitive behaviours, the Agent Behaviour Representation (ABR) formalism [37] is used. For reactive agent behaviours, AUML formalisms are used, specifically state charts. For interactions, protocol diagrams from AUML are used. | We used only adapted diagrams from AUML. For behaviours and knowledge representation, we employ Activity Diagrams. For interactions, we use Protocol Diagrams. |
| **Modelling process** | 1. Create a society of cognitive agents. Incorporate the informational flow.<br><br>2. Create a society of reactive agents. Incorporate the physical flow and the related exchanged objects (products).<br><br>3. Define the responsibility links between cognitive and reactive agents.<br><br>5. Specify agent behaviour of the cognitive society using the Agent Behaviour Representation (ABR) formalism.<br><br>6. Specify agent's behaviour of the reactive society using statecharts.<br><br>7. Specify agents' interactions through protocol diagrams. | Same process, but with different formalisms from AUML. |

The next sub-section provides some final remarks and conclusions about the proposed deployment strategy.

## 6. Final Remarks and Future Works

This paper presents a conversion strategy from the FAMASS analysis models into specification and design models inspired by the methodological agent-based framework of Labarthe et al. [9]. This strategy facilitates the FAMASS analysts in converting their models and going faster and smoother through the whole modelling process.

In addition, this deployment strategy demonstrates that the analysis phase of FAMASS can be integrated with other existing approaches specialized in specification and design modelling. With this as an impetus, other methodological frameworks could be inspected in the future so as to verify that FAMASS concepts adhere to other frameworks.

Furthermore, the proposed strategy allows us to avoid the research effort needed to develop a totally new specification and design methodology for the domain, although it would be suitable and desirable for future research initiatives. With regard to this, a forthcoming research effort will work on extending the FAMASS analysis approach, so as to cover the whole FAMASS life-cycle from analysis to simulation. In this way the proposed deploying strategy launches the basis for this FAMASS-extended version of a complete architecture to deal with agent-based simulations in the context of distributed APS systems. Future versions of the FAMASS approach are to be published shortly.

**Luis Antonio de Santa-Eulalia** is a professor at Téluq-UQAM (Université du Québec à Montréal), Canada, a member of the innovation board of Axia Value Chain (North America division), and a researcher of the NSERC Strategic Research Network on Value Chain Optimization (VCO). He holds a Ph.D. in Industrial Engineering from Université Laval, Canada, an MSc. and BSc. both in Industrial Engineering respectively from the University of São Paulo, Brazil, and Federal University of São Carlos, Brazil. He has worked as a researcher and consultant in the domains of production planning and control, supply chain management, and simulations. His current research interests are related to novel business models and technology for sustainable value chain management.

**Sophie D'Amours** holds a Ph.D. in Applied Mathematics and Industrial Engineering from the École Polytechnique de Montréal, as well as a MBA and a BSc in Mechanical Engineering from Université Laval. She is currently a scientific director of the NSERC Strategic Research Network on Value Chain Optimization (VCO) and she holds a Canada Research Chair in Planning Sustainable Forest Value Networks as well as an NSERC Industrial Chair. She is also Director of the FORAC Research Consortium. Sophie is a full professor at the Faculty of Science and Engineering, Department of Mechanical Engineering, at Université Laval. Her research interests are in supply chain management and planning, web-based applications, and forest sector.

**Jean-Marc Frayret** is Associate Professor at the École Polytechnique de Montréal, Québec, Canada. He holds a Ph.D. in Mechanical and Industrial Engineering from Université Laval, Canada. He is a member of the CIRRELT, a research centre dedicated to the study of network organizations and logistics. He is also a researcher of the NSERC Strategic Research Network on Value Chain Optimization (VCO) and of the FORAC Research Consortium. His research interests include agent-based and distributed manufacturing systems, supply chain management and interfirm collaboration. Dr. Frayret has published several articles in these fields in various journals and international conferences.